\documentclass[conference,9pt]{IEEEtran}
\usepackage{multirow}
\usepackage{booktabs}
\usepackage{graphicx}

\IEEEoverridecommandlockouts

\usepackage{multirow}
\usepackage{booktabs}
\usepackage{graphicx}
\usepackage{cite}
\usepackage{amsmath,amssymb,amsfonts}
\usepackage{algorithmic}
\usepackage{graphicx}
\usepackage{textcomp}
\usepackage{xcolor}
\usepackage{hyperref}

\usepackage{graphicx, tikz}
\usepackage{color}
\usepackage{kotex}
\usepackage{float}
\usepackage[normalem]{ulem}
\usepackage{multirow}
\useunder{\uline}{\ul}{}
\usepackage{hyperref}
\usepackage{booktabs}
\usepackage{subcaption}
\usepackage{amsfonts}
\usepackage{lineno}
\usepackage{url}
\usepackage{xparse}
\usepackage{musicography}

\def\BibTeX{{\rm B\kern-.05em{\sc i\kern-.025em b}\kern-.08em
    T\kern-.1667em\lower.7ex\hbox{E}\kern-.125emX}}

\begin{document}

\title{TokenSynth: A Token-based Neural Synthesizer \\ for Instrument Cloning and Text-to-Instrument
    \thanks{This work was partly supported by the National Research Foundation of Korea (NRF) grant funded by the Korea government (MSIT) [No. RS-2023-00219429, 50\%], Institute of Information \& communications Technology Planning \& Evaluation (IITP) grant funded by the Korea government(MSIT) [No. RS-2022-II220320, 2022-0-00320, Artificial intelligence research about cross-modal dialogue modeling for one-on-one multi-modal interactions, 40\%], [No. RS-2021-II212068, Artificial Intelligence Innovation Hub (Artificial Intelligence Institute, Seoul National University), 5\%], and [NO.RS-2021-II211343, Artificial Intelligence Graduate School Program (Seoul National University), 5\%]
    }
}


\author{
    \IEEEauthorblockN{
        Kyungsu Kim$^{\musSharp}$ \quad
        Junghyun Koo$^{\musSharp}$ \quad
        Sungho Lee$^{\musSharp}$ \quad
        Haesun Joung$^{\musSharp}$ \quad
        Kyogu Lee$^{\musSharp \musNatural \musFlat}$
    }
    \vspace{0.5cm}
    \IEEEauthorblockA{
        $\musSharp$\enspace Music and Audio Research Group (MARG), Department of Intelligence and Information, Seoul National University\\
        $\musNatural$\enspace Interdisciplinary Program in Artificial Intelligence, Seoul National University\\
        $\musFlat$\enspace Artificial Intelligence Institute, Seoul National University\\
        \texttt{\{kyungsu.kim, dg22302, sh-lee, gotjs3841, kglee\}@snu.ac.kr}
    }
}

\maketitle

\begin{abstract}
Recent advancements in neural audio codecs have enabled the use of tokenized audio representations in various audio generation tasks, such as text-to-speech, text-to-audio, and text-to-music generation. Leveraging this approach, we propose TokenSynth, a novel neural synthesizer that utilizes a decoder-only transformer to generate desired audio tokens from MIDI tokens and CLAP (Contrastive Language-Audio Pretraining) embedding, which has timbre-related information.
Our model is capable of performing instrument cloning, text-to-instrument synthesis, and text-guided timbre manipulation without any fine-tuning.
This flexibility enables diverse sound design and intuitive timbre control.
We evaluated the quality of the synthesized audio, the timbral similarity between synthesized and target audio/text, and synthesis accuracy (i.e., how accurately it follows the input MIDI) using objective measures.TokenSynth demonstrates the potential of leveraging advanced neural audio codecs and transformers to create powerful and versatile neural synthesizers. The source code, model weights, and audio demos are available at:
\href{https://github.com/KyungsuKim42/tokensynth}{\texttt{https://github.com/KyungsuKim42/tokensynth}}
\end{abstract}


\section{Introduction}

\begin{figure*}[t!]
    \centering
    \begin{subfigure}[t]{0.95\columnwidth}
        \centering
        \includegraphics[width=.955\linewidth]{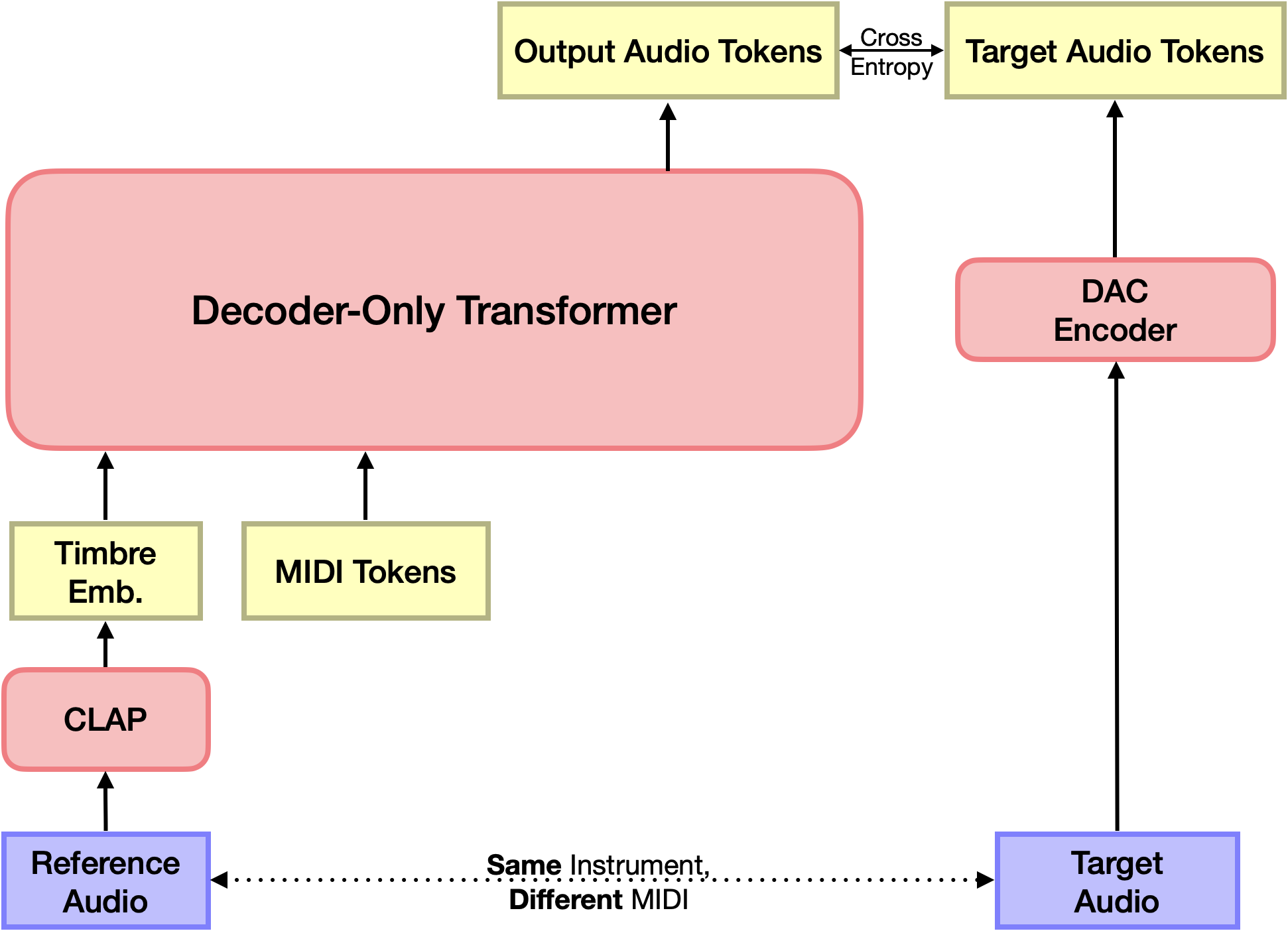}
        \caption{Training Phase}
    \end{subfigure}%
    \hspace{3pt}
    \tikz{\draw[densely dashed, thin](0,6cm) -- (0,0);}
    \hspace{3pt}
    \begin{subfigure}[t]{0.95\columnwidth}
        \centering
        \includegraphics[width=1.045\linewidth]{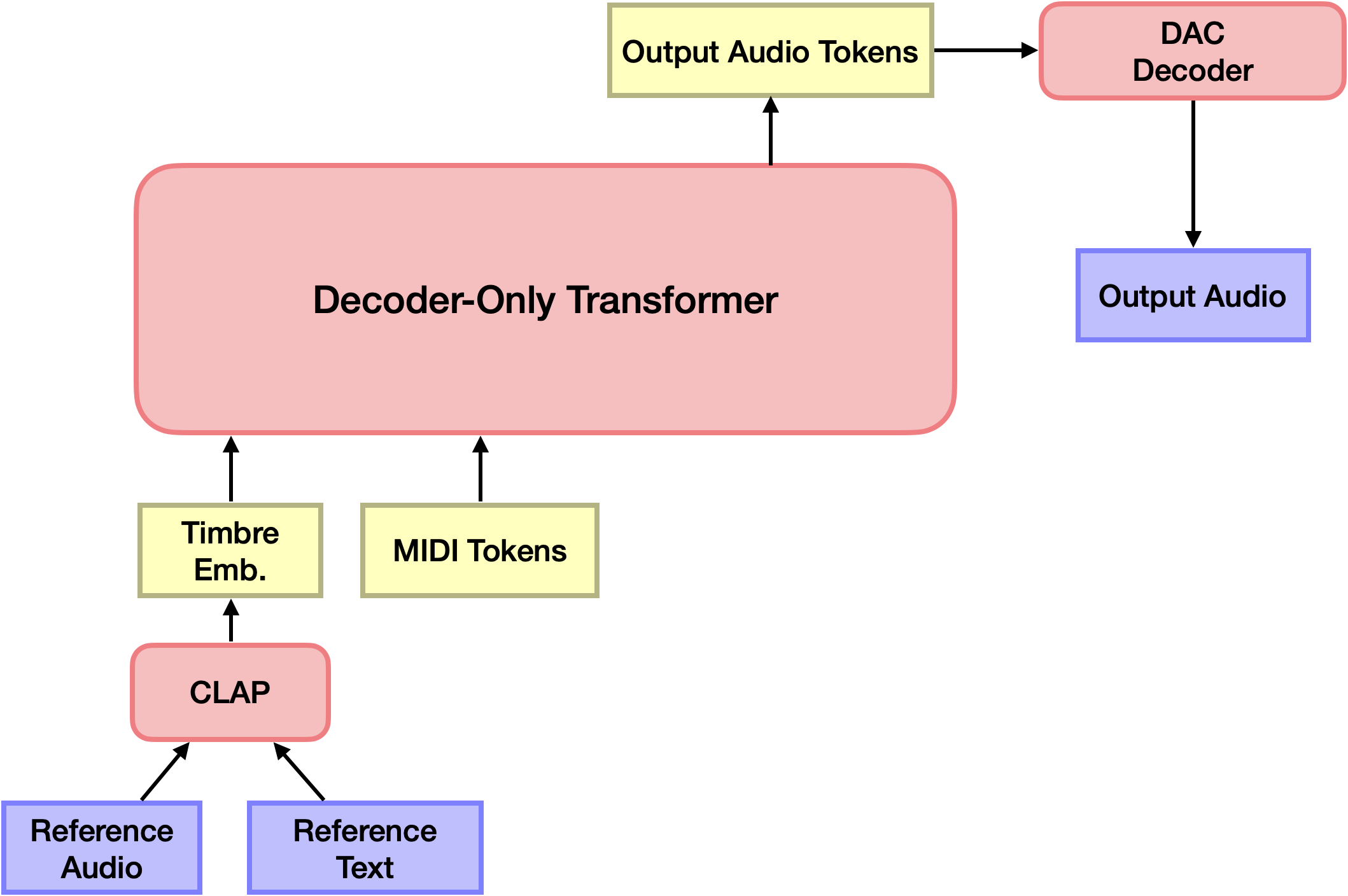}
        \caption{Inference}
    \end{subfigure}
    \caption{(a) TokenSynth takes a timbre embedding and MIDI tokens as input. The timbre embedding is extracted from reference audio using a pre-trained CLAP encoder. TokenSynth predicts the encoded audio tokens of the target audio using a pre-trained DAC encoder. (b) During inference, the timbre embedding is obtained from reference audio, text, or their combination, and the predicted tokens are decoded into audio using a DAC decoder.}
    \label{fig:overall}
\end{figure*}

Recent advancements in generative models, particularly in the audio domain, have led to remarkable progress in tasks such as text-to-speech, text-to-audio, and text-to-music synthesis~\cite{vall-e,audiogen,musicgen,musiclm,audioldm,diffwave}. These models are predominantly based on diffusion models~\cite{diffusion} and transformer architectures~\cite{transformer}, with the latter being well-suited for real-time audio tasks due to their autoregressive nature. The development of neural audio codecs~\cite{soundstream,encodec,dac}, which compress audio into a sequence of discrete tokens with a significantly lower frame rate than the original audio sample rate and reconstruct high-quality audio from these tokens, has been instrumental in enabling the use of transformers for audio generation~\cite{audiogen,musiclm,musicgen,vall-e}.

While music generation models have made significant strides~\cite{musicgen,musiccontrolnet,koosmitin}, they still face limitations in terms of user controllability when generating fully mixed music. In contrast, the real-world music creation process often involves composing multiple tracks using different instruments, then mixing them to create the final piece. This highlights the importance of considering instruments as essential building blocks in music production.

Using instruments as basic compositional units aligns with the natural music creation process.
In this context, efforts to expand the capabilities of instruments using deep learning have led to the development of new methods for controlling timbre~\cite{nsynthwavenet,gansynth,ddsp,sing,flexible,diffwaveshaping,diffwavetable,ddx7,ddspreview,fewsampleddsp,narita2023ganstrument}. Two prominent approaches are instrument cloning, which involves extracting timbres from existing musical audio~\cite{ddsp,fewsampleddsp}, and the recently proposed text-to-instrument synthesis~\cite{instrumentgen, sample-based}. Instrument cloning has been explored since the introduction of differentiable digital signal processing (DDSP)~\cite{ddsp,fewsampleddsp}, enabling the transfer of timbres from reference audio to synthesized sounds.

Existing instrument cloning methods often require fine-tuning and are limited to monophonic audio. 
Text-to-instrument models offer timbre control via natural language but rely on sample-based approaches that generate single-note samples, limiting expressive interactions in polyphonic instruments. An effective solution is an end-to-end polyphonic synthesizer that directly generates complex and expressive music with richer timbral control.

In this work, we propose TokenSynth, a polyphonic neural synthesizer capable of performing zero-shot instrument cloning and text-to-instrument simultaneously by leveraging neural codec language modeling. This marks the first use of an audio codec language model for a polyphonic neural synthesizer. We train a decoder-only transformer to generate target audio tokens given timbre embedding and MIDI tokens, using pretrained CLAP and neural audio codec models for timbre conditioning and audio tokenization, respectively. 
We generate a dataset of 9.53M samples by combining the NSynth and Lakh MIDI datasets~\cite{lakh}, following the approach proposed in Kim et al.~\cite{showme}, and further augment the dataset to double its size. 
During training, we only use audio to extract the CLAP embeddings for timbre conditioning, but at inference time, the model is able to utilize both audio and text as conditions, enabled by the CLAP encoder’s cross-modal representation learning~\cite{clap}, allowing for text-to-instrument and text-guided timbre manipulation through interpolation.

We train audio-to-MIDI transcription model and unconditional audio generation model to evaluate synthesis accuracy and apply classifier-free guidance~\cite{cfg, audiogen}.
We introduce a novel technique called First-Note Guidance, which applies classifier-free guidance only at the onset timing of the first note to stabilize the synthesis process. We conduct objective evaluations to assess the audio quality, timbral similarity, and synthesis accuracy of TokenSynth.

\section{Related Work}\label{sec:related_work}

\subsection{Neural Synthesizer}
Numerous attempts have been made to implement synthesizers using neural networks, primarily focusing on synthesizing the sounds of predefined instruments~\cite{sing} or enabling limited timbre control by interpolating between their timbres~\cite{nsynthwavenet,gansynth,flexible}.

DDSP~\cite{ddsp,ddspreview} marked a significant milestone in neural synthesizers by integrating digital signal processing components, effectively leveraging inductive biases from musical instruments to create lightweight models. Additionally, DDSP enabled instrument cloning, inspiring many subsequent works that adopted methodologies from digital synthesizers~\cite{diffwaveshaping,diffwavetable,ddx7}.

The recently proposed InstrumentGen~\cite{instrumentgen, sample-based} has made a significant contribution on text-to-instrument task by effectively leveraging neural codecs and transformers.
Our work follows a similar approach but aims to extend these capabilities beyond single-note audio generation to support instrument cloning from diverse audio recordings. 

\subsection{Neural Codec Language Modeling}

Neural audio codecs are models trained to encode audio into a compact sequence of discrete tokens and decode these tokens back into perceptually similar audio~\cite{soundstream,encodec,dac}. These models typically employ a VQ-VAE architecture~\cite{van2017neural} with residual vector quantization~\cite{rvq1,rvq2}, enabling efficient and high-quality audio representation. Operating as a lossy compression algorithm, neural audio codecs achieve significantly better audio quality at lower bitrates compared to traditional codecs like \texttt{mp3}~\cite{mp3}.

By learning high-level discrete audio representations, neural audio codecs enable transformers to autoregressively generate audio tokens via next-token prediction. These tokens are then decoded by the neural audio codec to produce the final audio output. Conditioning the transformer on paired text and target audio allows for text-conditioned audio generation. Models such as MusicLM~\cite{musiclm}, AudioGen~\cite{audiogen}, MusicGen~\cite{musicgen}, and VALL-E~\cite{vall-e} have successfully adopted this approach for tasks like text-to-speech, text-to-audio, and text-to-music synthesis.

\section{Method}

TokenSynth is a neural codec language model that autoregressively generates audio tokens $a \in \{1, …, K_a\}^{N\times D}$ conditioned on a timbre embedding $e \in \mathbb{R}^{d_{\text{CLAP}}}$ and MIDI tokens $m \in \{1, …, K_m\}^{M}$. Fig.~\ref{fig:overall} illustrates the overall architecture of TokenSynth during both the training and inference phases. We first explain how we extract the audio, MIDI, and timbre representations, then describe the language model training process based on these representations, and finally discuss how we perform audio synthesis using the trained transformer.

\subsection{Representations}
\subsubsection{Timbre Embedding}
We utilize a pretrained CLAP model~\cite{clap}, a cross-modal representation learning model inspired by CLIP~\cite{clip}, to obtain timbre embeddings. Trained on a large number of audio-text pairs, CLAP can generate embeddings from both audio and text inputs, mapping them into a shared embedding space. This shared representation enables TokenSynth to simultaneously perform instrument cloning, text-to-instrument synthesis, and text-guided timbre manipulation.

To evaluate the pretrained CLAP model’s ability to capture rich timbre information, we trained an MLP classifier on the extracted embeddings to classify 953 instruments, achieving a 90.4\% top-1 validation accuracy. This result confirms that CLAP features are sufficient for conditioning instrument timbre.

Since the dimensions of the CLAP encoder and token embeddings differ, we introduce a projection layer $\text{proj}(\cdot)$ using a 2-layer MLP and define $\hat{e} = \text{proj}(e) \in \mathbb{R}^{d_{\text{emb}}}$ as the timbre embedding.

\subsubsection{MIDI Tokenization}
MT3~\cite{mt3} pioneered the tokenization of MIDI data for transformer-based processing, and subsequent studies have adopted this approach. We also adopt this MIDI tokenization method with minor modifications.

Each note is represented using four token types:
\begin{itemize}
    \setlength\itemsep{-0.25em}
        \item Absolute onset timing (500 values): measured in 10 ms units
        \item Absolute offset timing (500 values): measured in 10 ms units
        \item Pitch (128 values): corresponding to 128 MIDI pitches
        \item Velocity (4 values): representing MIDI velocities
\end{itemize}

Therefore, a MIDI file with $n$ notes is represented by a sequence of tokens $m = m_{1:M} \in \{1,…,K_m\}^M$, where the sequence length is $M = 4n$.

\subsubsection{Audio Tokenization}

We adopt the Descript Audio Codec (DAC)~\cite{dac} as the neural audio codec for audio tokenization. DAC employs techniques such as periodic activation functions, complex STFT discriminators, and quantizer dropout to achieve higher compression rates while preserving audio quality. For single-channel audio $x \in [-1, 1]^T$, the DAC encoder converts it into quantized audio tokens $a = a_{1:N,1:D} \in \{1, …, K_a\}^{N \times D}$, where $D$ represents the codebook depth of the DAC quantizer obtained through residual vector quantization\cite{rvq1,rvq2}. The first depth code captures the most critical information. To effectively process audio tokens from multiple codebooks, we apply the delay pattern technique from MusicGen~\cite{musicgen}.

\subsection{Training}\label{subsection:training}
The model architecture and training method follow the standard decoder-only transformer approach except the model predicts $D$ number of tokens at each timestep. We use $D$ number of softmax layers to calculate the probability of tokens for each codebook. Training is performed by minimizing the cross-entropy loss for the next token prediction task, similar to a typical decoder-only transformer. The cross-entropy loss is calculated for each codebook as follows:

The model architecture and training method follow the standard decoder-only transformer approach, except that the model predicts $D$ tokens at each timestep. We use $D$ softmax layers to compute the token probabilities for each codebook. Training is performed by minimizing the cross-entropy loss for the next-token prediction task, as in a typical decoder-only transformer. The total cross-entropy loss across all codebooks is calculated as follows:

\begin{equation}
    \mathcal{L} = \sum_{n=1}^N\sum_{d=1}^{D}  -\log P_\theta(a_{n,d} | e,m,a_{1:n-1,1:D})
\end{equation}

To prevent the transformer from capturing performance-related information from the timbre embedding $e$, we ensure that the reference audio used to extract $e$ and the target audio to be generated share the same instrument but differ in performance.
This encourages the transformer to generate audio tokens that accurately represent the given MIDI sequence with the timbre encoded in the embedding.

Additionally, we train an unconditional audio generation model for applying classifier-free guidance during inference and an instrument-agnostic music transcription model to evaluate synthesis accuracy. 
The unconditional model generates audio tokens without timbre or MIDI conditioning, while the transcription model is trained without the timbre embedding and with the conditioning order of MIDI and audio tokens reversed.
Unlike MusicGen~\cite{musicgen}, these models are trained independently without parameter sharing with the main model.

\subsection{Inference}
During inference, we first extract the timbre embedding using the CLAP encoder. As CLAP is a cross-modal encoder, we can use either a reference audio, text, or multiple references and interpolate them. This interpolation between cross-modal representations enables text-guided timbre manipulation, allowing the modification of a reference audio's timbre using text.

Given the timbre embedding and MIDI tokens, we generate audio tokens using the top-p sampling method (nucleus sampling).

\subsubsection{First-Note Guidance}
To more strongly enforce the timbre condition, we optionally apply the classifier-free guidance method by extrapolating the logit values—inputs to the final softmax layer—as suggested in MusicGen~\cite{musicgen}.

However, applying classifier-free guidance at every time step can lead to unintended noise. Specifically, when both the conditional and unconditional models predict silent audio tokens, the guidance may inadvertently lower the logit value of the silence token, resulting in the sampling of unintended non-silent tokens.

To address this issue, we introduce the \textit{``first-note guidance”} technique, where classifier-free guidance is applied only at the time step $t’$ corresponding to the onset of the first note to be synthesized. Empirical results show that applying guidance solely at this onset is sufficient to control the timbre throughout the entire synthesized audio. This is because the training audio clips maintain a consistent timbre, enabling the model to generate audio with a stable timbre established at the first note.

\section{Experimental Setup}
\begin{table*}[t!]
\centering
\begin{tabular}{cccccc}
\hline
 & Ref=Tgt & Ref Dry/Wet & MSS Loss ($\downarrow$) & CLAP Score ($\uparrow$)& F-Score ($\uparrow$)\\ \hline
\multirow{2}{*}{Ground Truth} &  & Dry & - & 0.798 & 0.914 \\
 &  & Wet & - & 0.776 & 0.853 \\ \hline
\multirow{4}{*}{TokenSynth} & \multirow{2}{*}{True} & Dry & {\textbf{0.569}} & {\textbf{0.860}} & 0.643 \\
 &  & Wet & 0.747 & 0.804 & 0.645 \\ \cline{2-2}
 & \multirow{2}{*}{False} & Dry & 0.577 & 0.848 & 0.641 \\
 &  & Wet & 0.754 & 0.795 & 0.641 \\ \cline{1-2}
\multirow{4}{*}{TokenSynth-Aug} & \multirow{2}{*}{True} & Dry & 0.643 & 0.845 & {\textbf{0.837}} \\
 &  & Wet & 0.812 & 0.798 & 0.839 \\ \cline{2-2}
 & \multirow{2}{*}{False} & Dry & 0.666 & 0.833 & 0.832 \\
 &  & Wet & 0.826 & 0.790 & 0.832 \\ \hline
\end{tabular}
\caption{Objective evaluation on \textit{instrument cloning}.}
\label{table:cloning_objective}
\end{table*}

\subsection{Dataset}
To enable zero-shot instrument cloning on unseen instruments, the model must learn a general timbre representation by training on various instruments. Synthesizing arbitrary MIDI sequences also requires diverse MIDI data. We adopt the synthetic approach by Kim et al.~\cite{showme}, combining the NSynth and Lakh MIDI datasets to generate synthetic instrument audio. We rendered 10,000 five-second polyphonic audio samples per instrument in NSynth, producing 9.53 million MIDI-audio pairs for training and 530,000 pairs for testing.

To enhance timbre diversity, we augmented the audio using a digital effect chain with random parameters. Following Koo et al.~\cite{koo2023music}, we randomly applied EQ, distortion, and algorithmic reverb with a 0.5 probability, using parameters from predefined ranges. Identical effects were applied to each reference-target pair to maintain timbre consistency. This process doubled the dataset size.

\subsection{Model}
We used the pre-trained 44kHz version of DAC and CLAP with the checkpoint \texttt{\text{music\_audioset\_epoch} \texttt{\_15\_esc\_90.14.pt}}, freezing both models during training.

The decoder-only transformer has 12 layers, 16 heads, 1024 embedding, 4096 feed-forward dimensions, and 0.1 dropout.
The model has 175M parameters in total.

Models were trained on both the original and augmented datasets, referred to as TokenSynth and TokenSynth-Aug, respectively. Each model was trained for one epoch using the Adam optimizer~\cite{adam} with a learning rate of 1e-4, $\beta_1=0.9$, $\beta_2=0.999$, and a batch size of 8. This resulted in 1.2M steps for TokenSynth and 2.4M steps for TokenSynth-Aug. Mixed-precision training was used to reduce memory usage~\cite{torch-lightning,mixedprecision}.

\subsection{Evaluation Metrics}
We computed multi-scale spectral (MSS) loss, CLAP score, and F-score for objective evaluation. MSS loss measures the overall synthesis accuracy, considering both instrument cloning performance and adherence to the MIDI condition. We used the same parameters as suggested in DDSP to calculate MSS loss.

CLAP score measures timbral similarity between synthesized and target audio. For instrument cloning, it is computed as the cosine similarity between CLAP embeddings of the target and synthesized audio. In text-to-instrument tasks, the reference text is used to extract the CLAP embedding instead of the target audio, as there is no ground truth target audio.

F-score evaluates adherence to the MIDI condition. Synthesized audio is transcribed into MIDI note sequences using the audio-conditional MIDI transcription model described in Section~\ref{subsection:training}. The F-measure is calculated based on standard criteria for music transcription, including offset timing~\cite{mir_eval}.

\section{Results}
\begin{table}[h]
\centering
\begin{tabular}{lll}
\hline
 & CLAP Score ($\uparrow$) & F-Score ($\uparrow$) \\ \hline
TokenSynth &  \textbf{0.179} & 0.339 \\
TokenSynth-Aug & 0.159 & \textbf{0.8149} \\ \hline
\end{tabular}
\caption{Objective evaluation on \textit{Text-to-Instrument}.}
\label{table:tti}
\end{table}

\subsection{Instrument Cloning}
In this experiment, we synthesized audio by inputting 200 reference audio-MIDI pairs per instrument in the test set, using top-p sampling with $p=0.95$. This evaluated TokenSynth’s ability to clone the timbre of unseen instruments from short (5-second) audio samples and synthesize audio that matches the given MIDI. We compared TokenSynth, trained on a dry audio dataset, with TokenSynth-Aug, trained on an augmented dataset with digital audio effects.

The comparison covered cases where the reference and target audio either shared the same timbre and notes or had the same timbre but different notes. Results in Table 1 show slightly better performance when the reference and target audio were identical, though the difference was minimal. This suggests that even when extrapolating timbre between different pitch ranges, both models performed well.

The comparison included cases where the reference and target audio were identical or had the same timbre but different notes. Results in Table~\ref{table:cloning_objective} show slightly better performance when the reference and target audio were identical, though the difference was minimal. This suggests that both models perform well even when extrapolating timbre across different pitch ranges.

Surprisingly, TokenSynth, despite not being trained on wet audio, outperformed TokenSynth-Aug on wet audio across most metrics. Listening tests revealed that while TokenSynth-Aug tried to synthesize wet audio, its accuracy in applying audio effects was poor, likely due to the timbre embeddings from the CLAP model lacking information about audio effects.

TokenSynth outperformed TokenSynth-Aug in spectral loss and CLAP score, even surpassing the ground truth CLAP score. However, TokenSynth-Aug excelled in the F-score, which measures synthesis accuracy.

\subsection{Text-to-Instrument}

In the text-to-instrument experiment, we synthesized audio by sampling audio tokens using 10 text descriptions used in InstrumentGen~\cite{instrumentgen}, each paired with 200 distinct MIDI sequences as input. Top-$p$ sampling with $p=0.6$ was used, and first-note guidance with $\gamma=1.6$ was applied to strengthen the influence of text descriptions. Since there are no ground truth instrument-text pairs, we could not compute MSS loss and instead evaluated the models using CLAP score and F-score.

Table~\ref{table:tti} shows the CLAP score and F-score for TokenSynth and TokenSynth-Aug. The CLAP score represents the cosine similarity between the CLAP embeddings of the synthesized audio and the reference text. The CLAP scores for both models are much lower compared to the instrument cloning task, likely due to the modality gap problem~\cite{mindthegap}, where the cosine similarity between embeddings from different modalities is significantly smaller than that between embeddings from the same modality.

\subsection{Text-guided Timbre Manipulation}
Since the CLAP embeddings extracted from the reference audio ($e_a$) and reference text ($e_t$) reside in a shared embedding space, it is possible to interpolate between $e_a$ and $e_t$ to obtain an embedding $e_\alpha = \alpha \times e_t + (1-\alpha) \times e_a$ for timbre conditioning. This enables the manipulation of the reference audio's timbre in the direction of the text description, providing highly flexible timbre control beyond simply cloning the reference audio's timbre. Furthermore, it is possible to interpolate or even extrapolate between multiple CLAP embeddings.

Since the CLAP embeddings extracted from the reference audio ($e_a$) and reference text ($e_t$) share the same embedding space, they can be interpolated to obtain a timbre embedding $e_\alpha = \alpha \times e_t + (1-\alpha) \times e_a$. This allows flexible timbre manipulation, blending the reference audio’s timbre with the text description beyond simple timbre cloning. Additionally, interpolation between more than two embeddings or even extrapolation between multiple CLAP embeddings is possible.

As it is difficult to quantitatively evaluate the performance of this task we refer readers listen to the audio demos at \href{https://github.com/KyungsuKim42/tokensynth}{\texttt{https://github.com/KyungsuKim42/tokensynth}}.

\section{Limitations and Conclusion}

TokenSynth has several limitations. It cannot perform real-time synthesis as it requires the complete MIDI sequence before generating audio tokens. Additionally, due to the stochastic nature of autoregressive sampling, the model may not strictly follow the input MIDI notes. The MIDI tokenization also uses only four velocity values, limited by the NSynth dataset, resulting in less precise velocity control compared to other methods.

Despite these limitations, TokenSynth shows strong potential by enabling zero-shot instrument cloning, text-to-instrument synthesis, and text-guided timbre manipulation without fine-tuning. 
Our experiments validate its effectiveness in cloning timbres from unseen instruments and highlight the advantages of training with augmented data for improved performance and generalization across diverse timbres and audio conditions.

\bibliographystyle{ieeetr}
\bibliography{ref}

\end{document}